\documentclass[10pt,a4paper]{article}

\usepackage{times}

\usepackage{epsfig,wrapfig}
\usepackage{epstopdf} 

\usepackage{fancyhdr}

\usepackage{amsfonts}
\usepackage{subfig}

\usepackage[lmargin=2.5cm, rmargin=2.5cm,tmargin=3.50cm,bmargin=3.50cm]{geometry}
\usepackage{indentfirst}

\usepackage{titlesec}
\titleformat{\section}[hang]
  {\centering}{\thesection}{1ex}{\normalsize \textsc}
\titleformat{\subsection}[hang]
  {}{\thesubsection}{1ex}{\normalsize \textit}

\renewcommand{\thesection}{ \normalsize \textnormal{\Roman{section}.}}
\renewcommand{\thesubsection}{\normalsize \textnormal{\textsc{\textit{\Alph{subsection}.}}}}

\def\e{\begin{equation}}
\def\f{\end{equation}}
\def\_#1{{\bf #1}}

\def\.{\cdot}

\newcommand{\epsdyad}{\ensuremath{\overline{\overline{\epsilon}}}}
\newcommand{\mudyad}{\ensuremath{\overline{\overline{\mu}}}}
\newcommand{\Jdyad}{\ensuremath{\overline{\overline{J}}}}
\newcommand{\unitdyad}{\ensuremath{\overline{\overline{I}}}}

\begin{document}

\title{\large \textbf{Omega Transmission Lines}}
%
\def\affil#1{\begin{itemize} \item[] #1 \end{itemize}}
\author{\normalsize \bfseries \underline{J. Vehmas} and S. Tretyakov}
%
\date{}
\maketitle
\thispagestyle{fancy} 
\vspace{-6ex}
\affil{\begin{center}\normalsize Department of Radio Science and Engineering/SMARAD Center of Excellence, Aalto University, P.~O.~Box~13000, FI-00076 Aalto, Finland. \\
Email: joni.vehmas@aalto.fi
 \end{center}}

\begin{abstract}
\noindent \normalsize
\textbf{\textit{Abstract} \ \ -- \ \
In this paper, we study the relationship between omega media and periodically loaded transmission lines. It is shown under which conditions a periodically loaded transmission line can be treated as effective omega media. An example circuit is shown and analyzed.}
\end{abstract}

\section{Introduction}

Omega media is a widely researched reciprocal special case of bi-anisotropic media introduced in \cite{Saadoun}. The material relations in uniaxial omega media \cite{proposed} can be written as
\begin{equation}
\begin{array}{l}
\mathbf{D}=\epsdyad \cdot \mathbf{E} + j K \sqrt{\epsilon_0 \mu_0} \Jdyad \cdot \mathbf{H}
\\
\mathbf{B}=\mudyad \cdot \mathbf{H} + j K \sqrt{\epsilon_0 \mu_0} \Jdyad \cdot \mathbf{E},
\end{array} \label{eq:omegamatrel}
\end{equation}
where $K$ is the omega coefficient and $\Jdyad$ is the antisymmetric dyadic defined as $\Jdyad = \mathbf{z}_0 \times \unitdyad$. It has been suggested that omega media can be realized by embedding $\Omega$-shaped metal inclusions, i.e., centrally connected small dipole and loop antennas, into a conventional dielectric. However, as such wire omega particles are resonant structures with all the polarizabilities resonant always in the same frequency,
the operational bandwidth, i.e., the bandwidth where $K \neq 0$, and the tunability of medium parameters in general are very limited. 
This, in turn, limits the possibility for practical applications. Here, we consider possibilities of realizing omega media with periodically loaded transmission lines (TLs). First, we compare the wave impedance of omega media with the Bloch impedance of a general periodically loaded TL and derive required conditions for omega-like response. Second, a T-type circuit topology is considered to fulfill the required conditions.

\section{Omega media}

The propagation constant for axially propagating plane wave in omega media can be easily derived from (\ref{eq:omegamatrel}) and is given by \cite{proposed,Serdyukov}
\begin{equation}
\beta = k_0 \sqrt{\epsilon_t \mu_t - K^2} = k_0 \sqrt{\epsilon_t \mu_t}\sqrt{1 - K_n^2},
\label{eq:omegadispersion}
\end{equation}
where $\epsilon_t$ and $\mu_t$ are, respectively, the relative transverse permittivity and permeability and $K_n$ is the normalized omega coefficient defined as $K_n = K/\sqrt{\epsilon_t \mu_{t}}$. For lossless media with $\epsilon_t \mu_t > 0$, $K_n$ is purely real. Therefore, for such media the propagation constant is real, i.e., there is wave propagation, only when we have $|K_n| < 1$.
An interesting property that separates omega media from conventional magnetodielectric media is that the wave impedance is different for waves travelling in the opposite directions. The wave impedance for axial propagation can be written as \cite{proposed,Serdyukov}
\begin{equation}
Z_\Omega = \sqrt{\frac{\mu_0 \mu_t}{\epsilon_0 \epsilon_t}}\Big(\sqrt{1-K_n^2} \pm j K_n\Big),
\label{eq:omegaimp_normal}
\end{equation}
where the two solutions correspond to opposite axial propagation directions.


\section{Omega transmission lines}

\subsection{Required conditions for the TL unit cell}
Bloch impedance can be considered as the characteristic impedance of periodically loaded transmission lines. It is defined simply as the ratio of the voltage and current at the terminals of the unit cell. It should be noted that the value of the Bloch impedance depends on how the terminal points are chosen and is, therefore, not unique for a given unit cell. Bloch impedance for a general periodic structure is defined using ABCD-parameters as \cite{Pozar}
\begin{equation}
Z_B = \mp \frac{2 B}{A-D\mp\sqrt{(A+D)^2-4}}.
\label{eq:Bloch1}
\end{equation}
Here, the two signs correspond to different propagation directions and the current is defined to flow always in the direction of the energy propagation. It should be noted that the top sign does not necessarily always lead to the correct solution for the positively traveling wave and bottom sign for the negatively traveling wave but the solutions may switch. 
This can be seen as negative real part of the Bloch impedance. 
Taking this into account, the Bloch impedance for positively ($+$) and negatively ($-$) travelling waves can be written as
\begin{equation}
Z_{B\pm} = \frac{j B}{AD-1}\left(\sqrt{1-\left(\frac{A+D}{2}\right)^2}\pm j \frac{D-A}{2} \right).
\label{eq:Bloch3}
\end{equation}
Comparing the two impedances of (\ref{eq:Bloch3}) to characteristic impedances of omega media (\ref{eq:omegaimp_normal}) and assuming them to be equal, we can define the normalized omega coefficient $K_n$ using the Bloch impedances $Z_{B+}$ and $Z_{B-}$
\begin{equation}
K_n=\pm \frac{\frac{Z_{B+}-Z_{B-}}{Z_{B+}+Z_{B-}}}{\sqrt{1-\left(\frac{Z_{B+}-Z_{B-}}{Z_{B+}+Z_{B-}}\right)^2}}.
\label{eq:Kn_cond}
\end{equation}
This can be further written using ABCD parameters as 
\begin{equation}
K_n=\pm \frac{D-A}{2} \frac{1}{\sqrt{1-A D}}.
\label{eq:Kn_cond}
\end{equation}
Therefore, as long as we have $A \neq D$, i.e, the unit cell is asymmetric, the normalized omega coefficient $K_n$ is non-zero. It should be noted that here and in the following equations $\pm$ does not denote different propagation directions as before but simply different possible solutions. Furthermore, we can also determine the effective magnetodielectric wave impedance $\sqrt{\mu_t \mu_0/( \epsilon_t \epsilon_0)}$ based on (\ref{eq:omegaimp_normal}) and (\ref{eq:Bloch3}). This can be written as
\begin{equation}
\sqrt{\frac{\mu_t \mu_0}{\epsilon_t \epsilon_0}} = \pm \frac{j B}{\sqrt{1-A D}}.
\end{equation}
Knowing the effective normalized omega coefficient and the effective wave impedance, we can also extract the effective refractive index by comparing the dispersion in omega media (\ref{eq:omegadispersion}) and the dispersion in the periodically loaded TL. The latter can be calculated easily for any unit cell using basic ABCD-matrix theory and Floquet theorem and is given by \cite{Pozar}
\begin{equation}
\beta_\pm = \frac{1}{d} \ln \left(\frac{A+D\pm \sqrt{(A+D)^2-4(A D-B C)}}{2}\right),
\end{equation}
where $d$ is the period of the structure. For reciprocal unit cells, the two solutions are equal.

\subsection{Circuit topology}
\begin{figure}[h!]
\centering \epsfig{file=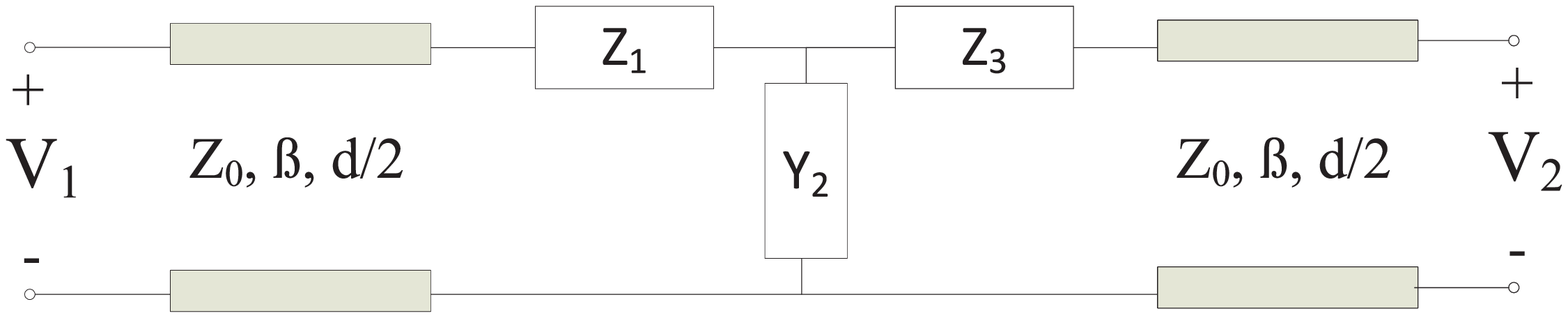, trim = .1cm .0cm .1cm .1cm, clip = true,width=0.65\textwidth}
\caption{Unit cell under study.} \label{fig:unitcell2}
\end{figure}

The simplest possible TL loading element is a T-type circuit (or alternatively $\pi$-type) as we need an asymmetric circuit for the Bloch impedances for different propagation directions to be different. Let us take a look at a TL loaded with a T-type circuit as shown in Fig.~\ref{fig:unitcell2}. Let us also assume that the the period of the unit cell $d$ is very small electrically. In this case, the ABCD-parameters, calculated by simply multiplying the ABCD matrices of each element in the unit cell in the right order, have the form
\begin{equation}
A = 1+ Y_2{Z_1}, \ \ \ B = Z_1 + Z_3 + Y_2 Z_1 Z_3, \ \ \ C = Y_2, \ \ \ D = 1+ Y_2{Z_3}.
\end{equation}
Therefore, the normalized omega coefficient, as defined in (7), can be written as
\begin{equation}
K_n = \pm  \frac{Y_2 (Z_3 - Z_1)}{\sqrt{1-(1+ Y_2{Z_1})(1+ Y_2{Z_3})}}
\end{equation}
and the magnetodielectric wave impedance as
\begin{equation}
\sqrt{\frac{\mu_t \mu_0}{\epsilon_t \epsilon_0}} = \pm \frac{j (Z_1 + Z_3 + Y_2 Z_1 Z_3)}{\sqrt{1-(1+ Y_2{Z_1})(1+ Y_2{Z_3})}}.
\end{equation}

Let us further simplify things by assuming that we have $Z_1=0 \ \Omega$. In this case, we have simply
\begin{equation}
K_n = \pm j \sqrt{ Y_2{Z_3}}, \ \ \ \ \ \ \ \sqrt{\frac{\mu_t \mu_0}{\epsilon_t \epsilon_0}} = \pm \sqrt{\frac{Z_3}{Y_2}}.
\end{equation}
From (13) rather surprisingly, it can be seen that even a TL periodically loaded with a simple series inductor - shunt capacitor circuit can be interpreted as omega media with $K_n = \pm \omega \sqrt{L C}$ and $\sqrt{\mu_t \mu_0 / (\epsilon_t \epsilon_0)} = \pm \sqrt{{L}/{C}}$. This is possible due to the asymmetric definition of the unit cell. If we define the terminal points of the unit cell so that the unit cell is symmetric (i.e., series-shunt-series loading with element values $Z_3/2$, $Y_2$ and $Z_3/2$), we have $D=A$ and thus, according to (7), $K_n = 0$ for all frequencies. Obviously, the dispersion in both cases should be the same as the physical structure is the same (assuming that $d$ is electrically small). On the other hand, this would mean that,  based on (2), the effective permittivity and permeability should be defined differently in the two cases!

In the presentation, this contradiction will be discussed in more detail and other circuit topologies will be analyzed.
\section{Conclusion}
We have discussed the connection between omega media and periodically loaded TLs. The effective omega material parameters for a general periodically loaded TL unit cell have been derived by comparing the characteristic impedances of omega media and a periodically loaded TL. An example of periodical loading by T-type circuit has been analyzed.



{\small

\begin{thebibliography}{10}
\setlength{\itemsep}{-1ex}

\bibitem{Saadoun}
M.M.I. Saadoun and N. Engheta, ``A reciprocal phase shifter using
novel pseudochiral or $\Omega$ medium," {\itshape Microwave and Opt. Tech. Lett.,} vol. 5,
pp. 184188, 1992.

\bibitem{proposed}
S.A. Tretyakov and A.A. Sochava, ``Proposed composite material for non-reflecting shields and antenna radomes,'' {\itshape Electronics Letters,} vol. 29, no. 12, pp. 1048-1049, 1993.

\bibitem{Serdyukov}
A. Serdyukov, I. Semchenko, S. Tretyakov and A. Sihvola, {\itshape Electromagnetics of Bi-anisotropic Materials,} Amsterdam, The Netherlands: Gordon and Breach Science Publishers, 2001.

\bibitem{Pozar}
D.M. Pozar, {\itshape Microwave Engineering,} 3rd edition, USA: John Wiley \& Sons, 2005.


\end{thebibliography}

}
\end{document}